\documentclass[aps,prb,groupedaddress,twocolumn,showpacs]{revtex4}
\usepackage{graphicx}

\begin{document}

\title{Field-induced superconductor to insulator transition in Josephson-junction ladders}

\author{Enzo Granato}

\address{Laborat\'orio Associado de Sensores e Materiais, \\
Instituto Nacional de Pesquisas Espaciais, \\
12245-970 S\~ao Jos\~e dos Campos, SP Brazil}

\begin{abstract}
The superconductor to insulator transition is studied in a
self-charging model for a ladder of Josephson-junctions in
presence of an external magnetic field. Path integral Monte Carlo
simulations of the equivalent ($1+1$)-dimensional classical model
are used to study the phase diagram and  critical behavior.  In
addition to a superconducting (vortex-free) phase, a vortex phase
can also occur for increasing magnetic field and small charging
energy. It is found that an intervening insulating phase separates
the superconducting from the vortex phases. Surprisingly, a
finite-size scaling analysis shows that the field-induced
superconducting to insulator transition is in the KT universality
class even tough the external field breaks time-reversal symmetry.
\end{abstract}

\maketitle
\newpage

Superconductor to insulator transitions in Josephson junction
arrays have attracted considerable interest
\cite{fazio,sondhi,geerligs,vdzant,kuo}. Such arrays can currently
be fabricated in any desired geometry both in one and two
dimensions \cite{vdzant,kuo} with well-controlled parameters. When
charging effects due to the small capacitance of the grains or
junctions dominate, strong quantum fluctuations of the phase of
the superconducting order parameter may drive the system into an
insulating phase at zero temperature leading to a superconductor
to insulator transition  as a function of charging energy or
external magnetic field. In two dimensions, the universality class
of these transitions have already been investigated in detail
numerically \cite{cha}, both in relation to experiments
\cite{geerligs,vdzant} and theoretical predictions
\cite{fisher,gk90,schon}. Nevertheless, there are also remarkable
quantum phase transition scenarios \cite{kardar} that can take
place in a Josephson-junction ladder in a magnetic field as in
Fig. 1 which have not been investigated experimentally or
numerically. Interestingly, a Josephson-junction ladder provides
the simplest one-dimensional version of the array allowing the
study of commensurability effects due to the flux lattice
\cite{kardar,eg90,eg92,giamarchi} or charge frustration
\cite{choi}, in presence of quantum fluctuations.

An important aspect of the Josephson-junction ladder in Fig. 1 is
the approximate relation to the quantum sine-Gordon model
\cite{haldane}, as first shown by Kardar \cite{kardar}. For
increasing magnetic field there is a transition from a (vortex
free) superconducting phase to a vortex phase where flux
penetrates the ladder, in absence of quantum fluctuations. This
transition is the analog of the commensurate-incommensurate
transition \cite{citrans} described by the sine-Gordon model. For
small fields, the phases in different branches of the ladder are
locked to each other while in the vortex state exponentially
interacting kinks (vortices) appear that unlock the phases leading
to a one-dimensional vortex lattice. Inclusion of quantum
fluctuations due to charging effects leads to the interesting
prediction that an insulating phase should occur in the vicinity
to this transition and therefore a direct transition from the
(vortex free) superconducting phase to the vortex lattice phase is
not possible even for small charging energies. So far, this
remarkable collective effect resulting from small capacitances in
the ladder has not been observed experimentally or even
numerically. The quantum critical behavior is also of great
interest. While at zero field the superconductor to insulator
transition is in the Kosterlitz-Thouless universality class
\cite{eg90,doniach}, the field-induced superconductor to insulator
transition could be in principle in a different universality class
since the magnetic field breaks the time reversal symmetry and the
commensurate-incommensurate transition in the sine-Gordon model is
in a distinct universality class \cite{haldane,citrans}.

In this work, we use path integral Monte Carlo (MC) simulations of
the equivalent ($1+1$)-dimensional classical model  to study the
phase diagram and  critical behavior of the Josephson-junction
ladder. The exchange MC method (parallel tempering) \cite{nemoto}
is used to determine more accurately the scaling behavior.  It is
found that an intervening insulating phase separates the
superconducting from the vortex phases, as shown in Fig. 2, in
agreement with Kardar's prediction \cite{kardar}. Thus, for
increasing field a single transition to an insulator occurs for
charging energies in the range $g_B < g < g_A$ while for lower
charging energies the superconductor to insulator transition is
followed by a transition into a vortex phase. Surprisingly, a
finite-size scaling analysis shows that the field-induced
superconducting to insulator transition AC is in the KT
universality class.

We consider a periodic Josephson-junction ladder as indicated in
Fig. 1, where  charging effects are  dominated by the capacitance
to ground of each grain \cite{doniach}, and described by the
Hamiltonian \cite{kardar}
\begin{equation}
{\cal H} = -{{E_c}\over 2} \sum_r \left( {d \over { d \theta_r }}
\right) ^2 - \sum_{<rr^\prime>} E_{rr^\prime} \cos ( \theta_r -
\theta_{r^\prime}-A_{rr^\prime}) \label{hamilt}
\end{equation}
The first term in Eq. (\ref{hamilt}) describes quantum
fluctuations induced by the charging energy  $E_c = 4 e^2 /C$ of a
non-neutral superconducting grain located at site $r$, where $e$
is the electronic charge and $C$ is the effective capacitance to
the ground of the grain, while  the second term is the usual
Josephson-junction coupling between nearest-neighbor grains
described by phase variables  $\theta_r$. The effect of the
applied magnetic field appears through the bond variables
$A_{rr^\prime}=(2\pi/\Phi_o)\int_r^{r'} A \cdot dr$, where $\bf A$
is the vector potential due to the external magnetic field $\bf B$
and the gauge-invariant sum around an elementary cell of the
ladder is given  by $\sum_{rr'} = 2 \pi f$ with $f=\Phi/\Phi_o$
defining the frustration parameter. It is sufficient to consider
$f$ in the range $[0,0.5]$.

\begin{figure}
\includegraphics[bb= 1cm 13cm  19cm   18cm, width=7.5 cm]{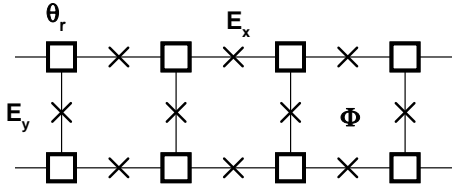}
\caption{ Schematic representation of a periodic
Josephson-junction ladder. Superconducting grains (squares) with
charging energy $E_c$ are coupled to the nearest neighbors  by
intra ($E_x$) and inter-chain ($E_y$) Josephson couplings
(crosses). The frustration $f$ corresponds to the magnetic flux
through the elementary cell, $f=\Phi/\Phi_o$, in units of the flux
quantum $\Phi_o$. }
\end{figure}

In order to  study the effects of the magnetic field  on the
superconductor to insulator transition and the critical behavior,
it is convenient to use an imaginary-time path-integral
formulation of the model \cite{sondhi}.  In this formulation, the
1D quantum problem of Eq. (1) maps into a 2D classical statistical
mechanics problem with the extra dimension corresponding to the
imaginary-time direction. The time axis $\tau$ can be  discretized
in slices $\Delta \tau$ and the ground state energy of the quantum
model corresponds to the reduced free energy of the classical
model, per unit length in the imaginary time direction. After
re-scaling the time slices appropriately in order to get
space-time isotropic couplings, the resulting classical partition
function is given by $Z = Tr_{\{\theta\}} e^{-H}$ where the
reduced classical Hamiltonian can be defined as
\begin{eqnarray}
H= &&-\frac{1}{g} \sum_{\tau,j} [
\cos(\theta_{\tau,j}-\theta_{\tau+1,j})
+\cos(\theta_{\tau,j}'-\theta_{\tau+1,j}') \cr &&
+\cos(\theta_{\tau,j}-\theta_{\tau,j+1}-\pi f) +
\cos(\theta_{\tau,j}'-\theta_{\tau,j+1}'+\pi f)  \cr &&+
{{E_y}\over{E_x}} \cos(\theta_{\tau,j} - \theta_{\tau,j}' )]
\label{chamilt}
\end{eqnarray}
In the above equation, $\theta$ and $\theta^\prime$ denote the
phases in the upper and lower branches in Fig. (1), $j$ and $\tau$
label the spatial and time directions, respectively, and the ratio
of the charging energy to the Josephson coupling $g =
(E_c/E_x)^{1/2}$ plays the role of an effective temperature in the
2D classical model. The equation (\ref{chamilt}) corresponds to a
gauge choice where the vector potential $A$ is parallel to ladder,
with opposite values in the upper and lower branches.

\begin{figure}
\includegraphics[bb= 1cm 9cm  19cm   21cm, width=7.5 cm]{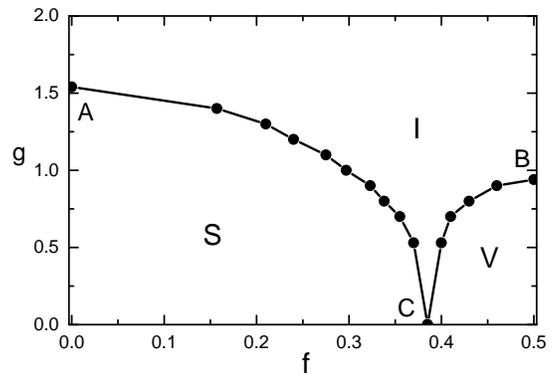}
\caption{Phase diagram of the ladder obtained from path integral
MC simulations showing the (vortex-free) superconducting phase
(S), the insulating phase (I) and the vortex phase (V), as a
function of frustration $f$ and ratio of charging to Josephson
energy $g = (E_c/E_x)^{1/2}$ for $E_y/E_x=2$. The critical field
$f_c$ at $g=0$ was obtained separately, by considering the ground
state of the ladder in absence of quantum fluctuations ($E_c=0$).
The lines are just guides to the eye.}
\end{figure}

We carry out MC simulations using the 2D classical Hamiltonian in
Eq. \ref{chamilt} regarding $g$ as a temperature-like parameter
and employing the standard Metropolis's algorithm to generate the
equilibrium distribution. Near the superconductor to insulator
transition, we use the exchange MC method (parallel tempering)
\cite{nemoto} to determine more accurately the scaling behavior of
various physical quantities. In fact, this method is known to
reduce significantly the critical slowing down near the
transition. In this method, many replicas of the system with
different couplings $g$ in a range above and below the critical
point are simulated simultaneously and the corresponding
configurations are allowed to be exchanged with a probability
distribution satisfying detailed balance. Simulations are
performed in system sizes with equal spatial and time linear
length $L$. This choice of the aspect ratio of the system assumes
implicitly that the dynamic critical exponent $z$ characterizing
the superconductor to insulator transition is close to $z \sim 1$.
In general, a quantum phase transition is characterized by
intrinsic anisotropic scaling with different diverging correlation
lengths $\xi$ and $\xi_\tau$ in the spatial and time directions
\cite{sondhi}, respectively, related by the dynamic exponent $z$
as $\xi_\tau \propto \xi^z$. Our choice is justified by the
analysis discussed below, showing that the scaling behavior is in
fact consistent with $z=1$.

\begin{figure}
\includegraphics[bb= 1cm 3cm  19cm   26cm, width=7.5 cm]{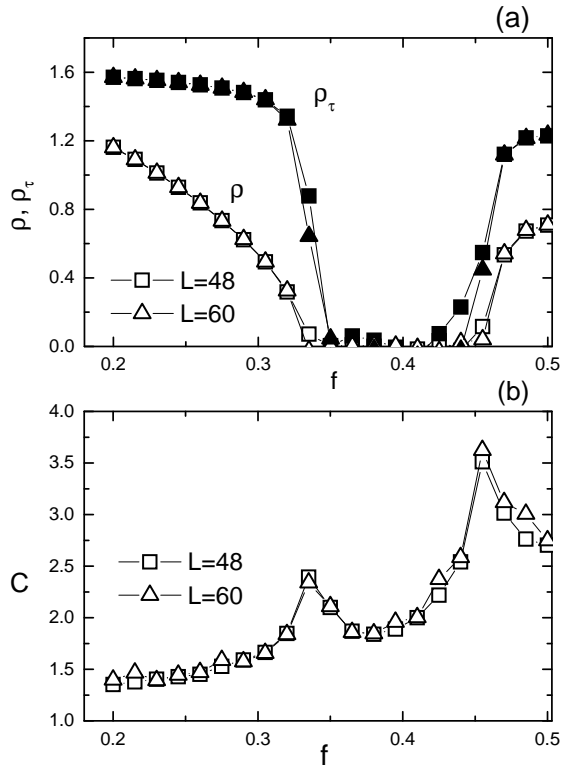}
\caption{ Phase stiffness in spatial $\rho$ and imaginary time
$\rho_\tau$ directions (a) and energy fluctuation $C=
(<H^2>-<H>^2)/L^2$ of the equivalent classical model (Eq.
(\ref{chamilt})), as a function of frustration $f$,  for a
constant value of the ratio $g=0.9$. The double peaks in $C$
indicate two transitions for increasing $f$ with an intervening
insulating phase corresponding to vanishing phase stiffness. }
\end{figure}

To locate the superconductor to insulator transition as a function
of charging energy and magnetic field, we have performed MC
calculations of the helicity modulus (phase stiffness), $\rho$ and
$\rho_\tau$,  in the spatial and time directions. In the
superconducting phase these quantities should be finite,
reflecting the existence of phase coherence, while in the
insulating phase they should vanish in the thermodynamic limit.
Fig. 3a shows the behavior of the helicity modulus as a function
of frustration for a fixed value of $g$, below the critical value
corresponding to $f=1/2$ frustration in Fig. 2, $g_B \sim 0.96$.
The helicity modulus remains finite at small fields and at fields
close to $f=1/2$ where a superconducting phase is expected
\cite{eg92} but at intermediate fields reach small values which
decreases with increasing system sizes. Fig. 3b also show the
behavior of the energy fluctuation (specific heat of the classical
model) $C$ as a function of frustration showing two peaks near the
two critical fields defining the region where the helicity modulus
reach small values. The behavior in Figs 3a and 3b  indicate that
there are two superconducting to insulator transitions for
increasing $f$ with an intervening insulating phase corresponding
to vanishing phase stiffness for large system sizes $L$.
Performing the same calculations  for different couplings $g$, we
have constructed the phase diagram in Fig. 2 for $E_y/E_x=2$
showing three different phases separated by the transition lines
AC and BC with the insulating phase extending between the
superconducting vortex-free phase S and vortex phase V for $g <
g_B$. The critical field $f_c$ at $g=0$ was obtained separately,
by considering the ground state of the ladder in absence of
quantum fluctuations ($E_c=0$). Using simulated annealing to
obtain the ground state, $f_c$ was determined as  the value of $f$
where flux first penetrates the ladder for increasing $f$.  Note
that, due to the long runs required to reach proper equilibration,
the lowest nonzero value of $g$ accessible within the present MC
calculations is $g \sim \frac{1}{2} g_B$. Therefore, this result
does not rule out the possibility that the two transitions lines
in Fig. 2 merge at a lower value of $g$ into a single line ending
at $f_c$ in which case a direct transition from S to V phases is
possible for sufficiently small $g$. We have performed additional
calculations with various ratios $E_y/E_x $ and found that the
insulating phase is more clearly visible for $E_y/E_x > 1.5$ while
at lower values, the transition lines AC and BC are difficult to
resolve at low values of $g$.

The intervening insulating phase between the superconducting phase
at low fields and vortex phase at higher fields was predicted
analytically in an earlier work by Kardar \cite{kardar} using an
approximate relation of the Hamiltonian of Eq. (\ref{hamilt}) with
Gaussian and sine-Gordon models describing fluctuations of the
phase variables $\phi = (\theta'+\theta)/2$ and
$\psi=(\theta'-\theta)/2$, respectively, where $\theta'$ and
$\theta$ denote the phases in the  upper and lower branches of the
ladder in Fig. (1). This approximation allows one to calculate the
power-law correlation functions for $\theta$ and $\theta'$, $C(r)
\propto 1/r^{\eta}$, describing the superconducting and vortex
phases, using the exact results for the sine-Gordon model
\cite{haldane}. However, the assumed Gaussian approximation leaves
out space-time vortices which are responsible for the insulating
phase where the correlation function decays exponentially. The
insulating phase was then studied by adding the effects of
space-time vortices "by hand" through the usual KT criterion $\eta
< 1/4$ to determine the bound of stability of the power-law
correlated phase. The results showed that the insulating phase
expected for large $g$, extends all the way to $g=0$ at $f_c$. The
phase diagram in Fig. 2 is consistent with this prediction but as
mentioned above a direct transition between S and V phases at
sufficiently small $g$ can not be ruled out within the present
calculations.

Although a KT-like stability criterion was used in Kardar's work
\cite{kardar} to determine approximately the  location of the
transitions to the insulating phase in the phase diagram, the
nature of the transition can not be determined by such argument
because it is based in decoupled Gaussian and sine-Gordon models.
Like in a similar analysis employing the KT criterion for the
sine-Gordon model \cite{haldane83}, applied to adsorbed layers,
which also predict an intervening disordered phase near the
commensurate-incommensurate transition for low-order commensurate
phases, the universality class of the commensurate to disordered
phase has to be determined by other methods based on symmetry
considerations, mapping to a solvable model or numerical
simulations.

\begin{figure}
\includegraphics[bb= 1cm 1cm  19cm   28cm, width=7.5 cm]{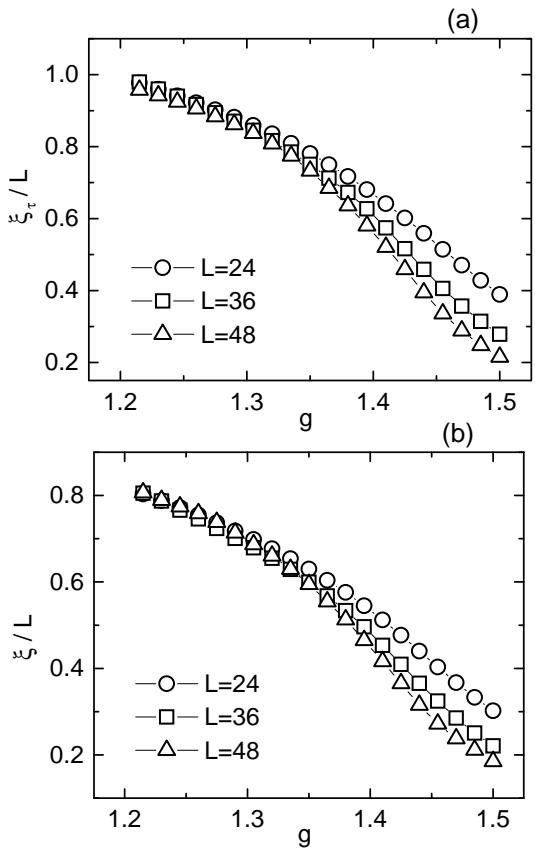}
\caption{ Behavior of the scaled correlation length $\xi_\tau/L$
(a) and $\xi/L$ (b) across the superconductor to insulator
transition line AC in Fig. 2 at a fixed nonzero frustration
$f=0.2$, for different systems sizes $L$.   }
\end{figure}

The universality class of the field-induced superconductor to
insulator transition along the line AC in Fig. 2 is of particular
interest. At zero magnetic field ($f=0$) the transition as a
function of charging energy is in the KT universality class
\cite{doniach,eg90} while at nonzero magnetic fields in principle
a different universality class is possible since the magnetic
field breaks the time-reversal symmetry. In addition, the relation
of the ladder model of Eq. (1) with the quantum sine-Gordon chain
\cite{kardar} also suggests the possibility of a universality
class different from the KT transition at zero field since the
superconductor to insulator transition in the ladder is driven by
the commensurate-incommensurate transition in the sine-Gordon
model which is in a distinct universality class
\cite{haldane,citrans}. In order to study the critical behavior of
the transition line AC in detail we have investigated the
finite-size behavior of the phase correlation lengths $\xi$ and
$\xi_\tau$ in addition to the helicity modulus $\rho$ and
$\rho_\tau$ near the transition line AC. The correlation lengths
$\xi(L)$ and $\xi_\tau(L)$ in the finite-size system can be
obtained from a second moment calculation using the correlation
function as \cite{cooper}
\begin{equation}
\xi(L) = \frac{1}{2\sin(k/2)}[S(0)/S(k) - 1]^{1/2}
\end{equation}
where $S(k)$ is given by the Fourier transform
\begin{equation}
S(k)=\sum_{x} <e^{i(\theta_{\tau,j} -\theta_{\tau,j+x}}> e^{i k x}
\end{equation}
The wavector $k=2\pi/L$ is the smallest value allowed in the
finite system. Similar expressions are used for the correlation
length $\xi(L)_\tau$ in the time direction.

Fig. 4 shows the finite-size behavior of the correlation length
scaled by the system size $L$  near the transition line AC in Fig.
2, as function of $g$, for a fixed value of frustration $f=0.2 <
f_C$. In the insulating phase for $g > g_c$, the correlation
length is finite and therefore $\xi/L$ and $\xi_\tau/L$ decrease
with system size and should vanish in the limit $L \rightarrow
\infty$ while at the transition and in the superconducting phase,
where the system is critical, $\xi/L$ should approach a finite
value depending of $g$. The behavior in Fig. 4a and 4b, showing
that both $\xi/L$ and $\xi_\tau/L$ merge for low values of $g$ at
approximately the same critical value $g_c\sim 1.3$ is consistent
with a transition in the KT universality class where the
correlation length diverges exponentially $\xi \propto e^{
b/|g-g_c|^{1/2} }$. In general, for a power-law correlation length
$\xi \propto |g-g_c|^{-1/2}$, the dimensionless quantity $\xi/L$
is expected to cross at the transition for different system sizes
\cite{cooper} and satisfy the scaling relation $\xi/L
=G((g-g_c)L^{1/\nu})$ near the transition. Then, the slopes $s(L)$
of the curves $\xi/L$ for different $L$ are determined by the
critical exponent $\nu$ as $s(L)= \frac{\partial}{\partial g}
(\xi/L) \propto L^{1/\nu}$, evaluated near $g_c$. The exponential
correlation length of the KT transition corresponds to $\nu
\rightarrow \infty$ which implies that $s(L)$ should be
$L$-independent and therefore the curves $\xi/L$ for different
system sizes should merge at the critical point, as in fact
observed in Fig. 4.

\begin{figure}
\includegraphics[bb= 1cm 1cm  19cm   28cm, width=7.5 cm]{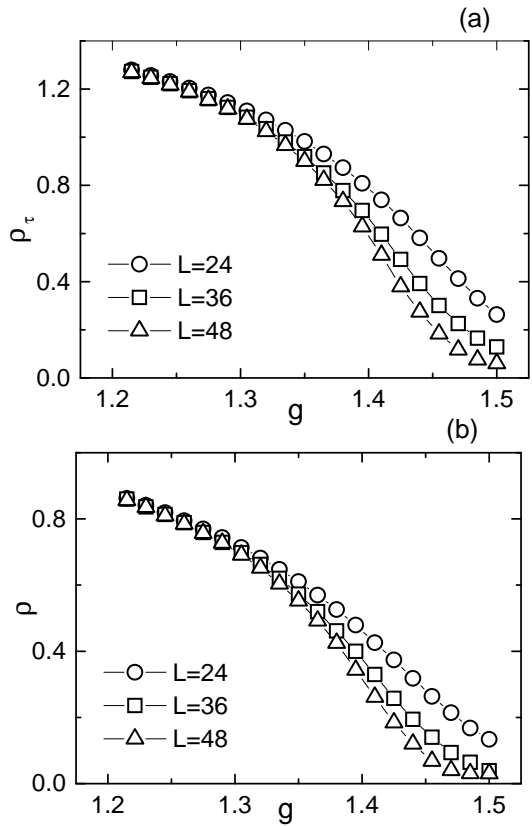}
\caption{ Behavior of the phase stiffness $\rho_\tau$ (a) and
$\rho$ (b) across the superconductor to insulator transition line
AC in Fig. 2 at a fixed nonzero frustration $f=0.2$, for different
systems sizes $L$.   }
\end{figure}

\begin{figure}
\includegraphics[bb= 1cm 6cm  19cm   23cm, width=7.5 cm]{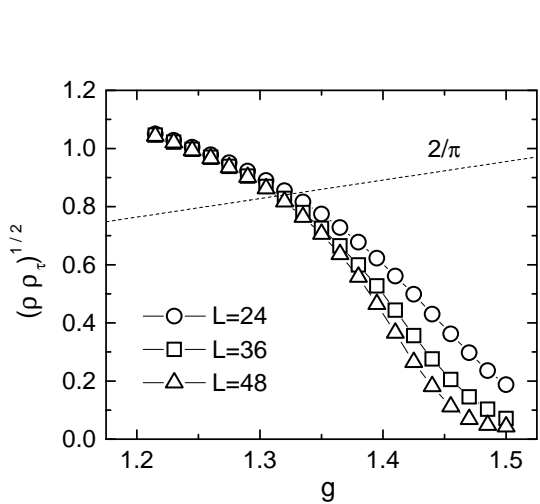}
\caption{ Behavior of the quadratic average of the phase stiffness
$(\rho \rho_\tau)^{1/2}$ across the superconductor to insulator
transition line AC in Fig. 2 at a fixed nonzero frustration
$f=0.2$, for different systems sizes $L$. Dotted line corresponds
to the universal jump prediction for a KT transition.}
\end{figure}

The finite-size behavior of the helicity modulus $\rho$ and
$\rho_\tau$ is also consistent with a KT transition. As shown in
Fig. 5, $\rho$ and $\rho_\tau$ also merge for decreasing $g$ at
approximately the same critical $g_c$. At the transition and in
the superconducting phase, where the system is critical, these
quantities  should scale as \cite{sondhi,cha} $\rho \propto
L^{2-d-z}$ and $\rho_\tau \propto L^{z-d}$, in $d$ spatial
dimensions. Since $\rho_\tau$ and $\rho$ in Fig. 5a and 5b
approach finite values independent of $L$ in this regime, the
dynamic exponent is $z=1$ for the present case where $d=1$. This
value of the dynamic exponent $z$ means that the line of Gaussian
fixed points can be made isotropic even though the phase
stiffness, $\rho$ and $\rho_\tau$, are different in the spatial
and time directions. In fact, by a suitable rescaling of the
coordinates in the spatial and time directions, the stiffness of
the renormalized isotropic Gaussian model can be written as $\bar
\rho=(\rho \ \rho_\tau)^{1/2}$. If the transition is in the KT
universality class then a universal jump \cite{nelson} is expected
for $\bar \rho$ as a function of $g$ at the transition,  given by
$\bar \rho = \frac{2}{\pi}g $. Fig. 6 shows  that the finite-size
behavior of $\bar \rho$ appears consistent with the expected
universal jump for the phase stiffness. For each system size, the
intersection of the data with the dotted line representing the
universal jump prediction provides an upper bound estimate of the
critical coupling. A more quantitative determination of the
critical coupling and the corresponding jump in the phase
stiffness can be obtained by the Weber and Minnhagen's scaling
analysis, which has been introduced before as an accurate method
of locating the critical coupling and the universal jump of the KT
transition for the ordinary XY model \cite{weber}. The phase
stiffness at the transition should have a logarithmic correction
given by \cite{weber}
\begin{equation}
\rho(L) = \rho_R( 1 + \frac{1}{2} \ \frac{1}{\ln L + C} )
\end{equation}
where $C$ is an undetermined constant and the universal jump
implies $\rho_R=\frac{2}{\pi} g_c$. Using data for different
system sizes, the best numerical fit according to this scaling
form performed for different values of the coupling $g$ can then
be used to locate the transition and determine the jump. Treating
both $\rho_R $ and $C$ as free parameters, we have performed this
fit using data of the quadratic averaged phase stiffness $\bar g$
for smaller system sizes $L \le 24$, where the data is more
accurate. As shown in Fig. 7, the value of the jump near the
location of the minimum in the fitting error for different
couplings $g$, is indeed consistent with the value expected for a
KT universality class.

\begin{figure}
\includegraphics[bb= 1cm 1cm 19cm 28cm, width=7.5cm]{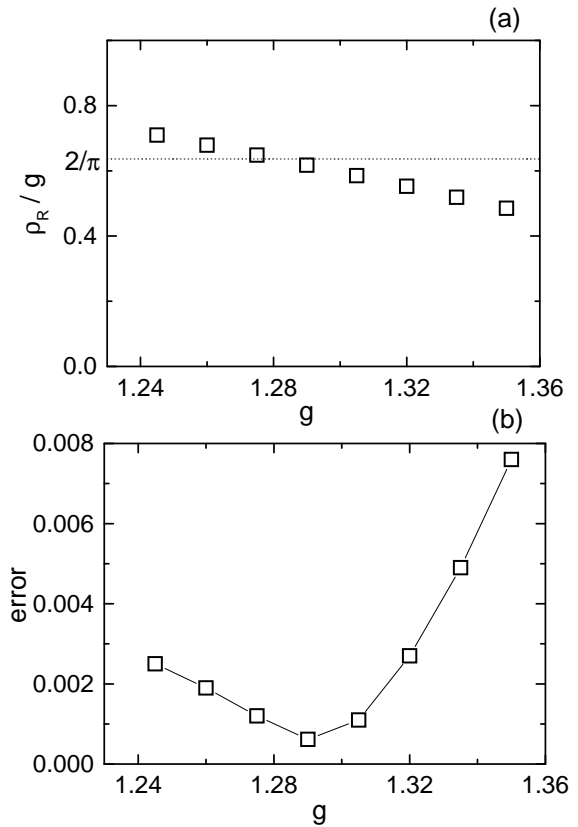}
\caption{ Numerical fit of the phase stiffness $\bar \rho$
according to Weber and Minnhagen's scaling relation with two free
parameters for system sizes $L=6,8,12,16,18,24$, at different
couplings $g$ across across the superconductor to insulator
transition. (a) The estimate of the jump $\rho_R/g_c$ from the
fitting parameter and (b) error estimate of the fit. The location
of the minimum in (b) determines the critical coupling $g_c$.
Dotted line in (a) is the value of the universal jump for a KT
transition. }
\end{figure}

In conclusion, we have studied the superconductor to insulator
transition in a ladder of Josephson-junctions under an applied
magnetic field by path integral MC methods. We found clear
numerical evidence in the phase diagram of an intervening
insulating phase between the superconducting and vortex phases, in
good agreement with an earlier prediction by Kardar \cite{kardar}.
In addition, we find that the field-induced superconducting to
insulator transition is in the KT universality class. In the
vortex phase, commensurability of the vortex lattice and the
ladder \cite{eg90,giamarchi} will strongly depend on the
frustration parameter $f$. Since Josephson-junction arrays can
currently be fabricated in any desired geometry and with
well-controlled parameters, our numerical evidence of the
intermediate insulating phase and the universality class of the
transition should be interesting enough to motivate experiments in
these intriguing systems.

\medskip

This work was supported  by Funda\c c\~ao de Amparo \`a Pesquisa
do Estado de S\~ao Paulo (FAPESP, Proc. no. 03/00541-0).

\end{document}